\documentclass[conference]{IEEEtran}
\IEEEoverridecommandlockouts
\usepackage{amsmath,amssymb}
\usepackage{booktabs}
\usepackage{graphicx}
\usepackage{xcolor}
\usepackage{pgfplots}
\usetikzlibrary{arrows.meta,positioning,fit,backgrounds,calc}
\pgfplotsset{compat=1.18}
\usepackage[hidelinks]{hyperref}
\usepackage{cleveref}
\usepackage{eso-pic}
\AddToShipoutPictureBG*{\AtPageLowerLeft{\put(36,18){\parbox{540pt}{\fontsize{6}{7}\selectfont
\copyright\ 2026 IEEE. Personal use of this material is permitted. Permission from IEEE must be obtained for all other uses, including reprinting or republishing this material for advertising or promotional purposes, creating new collective works, for resale or redistribution to servers or lists, or reuse of any copyrighted component of this work in other works.}}}}

\begin{document}
\title{Measuring Consensus in AI Agent Networks}
\author{\IEEEauthorblockN{Jin Gao}
\IEEEauthorblockA{Independent Researcher, USA\\gaojin@alum.mit.edu}}
\maketitle

\begin{abstract}
As AI agents become increasingly connected through networks and take on tasks that require
negotiation and joint decision-making, a systematic approach is needed to assess how their
views converge and what information supports their decisions. We present an evaluation
framework that treats consensus formation as the evolution of publicly stated positions
toward a shared region, using accumulated communication history as context. The framework
tracks position trajectories, group dispersion, and information expressed in public messages.
Two case studies examine staffing decisions based on distributed facts and decisions
requiring agents to reconcile duplicate information and a cancelled contract. Results from 1,872 completed runs
associate sharing explanations with more accurate group decisions, while the second task
reveals persistent disagreement despite accurate aggregate answers. Membership-change
experiments provide a complementary evaluation. By separating convergence among agents from
progress toward a task's full-information answer, the framework relates decision accuracy
and bounded disagreement to the information available during the exchange.
\end{abstract}


\section{Introduction}
Networked AI agents exchange information and proposals to address shared problems, such as planning
staffing from facts held by different participants. As the discussion develops, their stated positions
may move closer together, remain divided, or change direction when new information is shared.
Evaluating this process calls for tracking both the evolution of group positions and the information
expressed during the exchange. Research on agreement protocols~\cite{dolev1986,mendes2013multidim,vaidya2013vector},
LLM-agent coordination~\cite{jo2025decentllms}, and distributed information use~\cite{li2025hiddenbench}
provides complementary perspectives on these questions. We bring these perspectives together in an
evaluation framework that follows consensus formation through the accumulated history of public communication.

Classical agreement protocols coordinate distributed processes through rule-based updates to scalar or vector
values~\cite{dolev1986,vaidya2013vector}; applications of approximate agreement include clock synchronization
and stabilization of sensor readings~\cite{dolev1986}. We study networks of autonomous LLM agents that generate
proposals and revise their stated positions through language-based interaction. Our focus is on consensus
states and the process through which they form. A consensus state describes publicly stated positions
occupying a shared region within a task-defined tolerance. Consensus formation describes how these positions
evolve through successive exchanges, interpreted in the context of accumulated public communication.
The framework tracks both the distribution of positions and their evolution alongside the information
expressed in the exchange. In our case studies, a full-information answer provides a separate benchmark
for evaluating the group's outcome.

This paper presents an evaluation framework for describing consensus states and tracing their formation
through publicly expressed positions and information. We examine the framework
through two case studies of distributed-information decisions, complemented by membership-change
experiments with LLM agents deployed on a Docker network.
Together, these studies connect the measurement of group convergence with the information exchanged
during deliberation.

\section{Related Work}
\textbf{(1) Distributed information.} Studies examine how groups use information held by different
participants, including HiddenBench~\cite{li2025hiddenbench}, SILO-BENCH~\cite{zhang2026silobench},
and InfoDelphi~\cite{li2026infodelphi}. \textbf{(2) Opinion dynamics.} Studies examine how positions
and claims evolve during interaction, including Hao et al.~\cite{hao2026flips} on stance convergence,
DEBATE~\cite{chuang2025debate} on opinion dynamics, and TIDE~\cite{yan2026tide} on misinformation
propagation. \textbf{(3) Numerical agreement.} Studies formalize agreement and validity conditions
for scalar or vector values, including Dolev et al.~\cite{dolev1986} on approximate agreement,
Mendes and Herlihy~\cite{mendes2013multidim} on multidimensional agreement, and Vaidya and
Garg~\cite{vaidya2013vector} on vector consensus. Our evaluation focuses on consensus states and
their formation in networked LLM agents, relating publicly stated positions to accumulated communication history.

\section{Evaluation Framework}\label{sec:framework}
\begin{figure*}[t]
\centering
\begin{tikzpicture}[scale=1.0, font=\scriptsize, line width=0.4pt,
  box/.style={draw=black!70, rounded corners=2pt, fill=white, align=center, inner sep=3pt},
  pane/.style={draw=black!45, rounded corners=3pt, fill=black!3, inner sep=6pt},
  ag/.style={draw=black!70, circle, fill=white, minimum size=10pt, inner sep=0pt, font=\tiny},
  lbl/.style={font=\scriptsize\bfseries, black!75},
  wire/.style={draw=black!55, -{Stealth[length=3pt]}},
  faint/.style={draw=black!25},
  cut/.style={draw=black!45, dashed, -{Stealth[length=3pt]}}]
\node[ag] (n1) at (0.15,1.75) {};   \node[ag] (n2) at (1.15,2.15) {};
\node[ag] (n3) at (2.05,1.55) {};   \node[ag] (n4) at (0.55,0.15) {};
\node[ag] (n5) at (1.95,-0.2) {};   \node[ag] (n6) at (-0.2,0.85) {};
\foreach \a/\b in {n1/n2,n2/n3,n1/n6,n6/n4,n4/n5,n5/n3,n2/n6}{\draw[faint] (\a)--(\b);}
\node[ag] (a0) at (0.75,1.15) {$a_0$};  \node[ag] (a1) at (1.75,0.95) {$a_1$};
\node[ag] (a2) at (1.25,0.45) {$a_2$};  \node[ag] (a3) at (0.45,0.62) {$a_3$};
\foreach \a/\b in {a0/a1,a0/a2,a0/a3,a1/a2,a1/a3,a2/a3}{\draw[black!60] (\a)--(\b);}
\draw[black!35, dashed] (a3) -- (n6);
\begin{scope}[on background layer]
  \node[pane, fit={(a0)(a1)(a2)(a3)}, inner sep=7pt] (scope) {};
  \node[pane, fill=none, draw=black!30, fit={(n1)(n2)(n3)(n4)(n5)(n6)(scope)}, inner sep=8pt] (net) {};
\end{scope}
\node[lbl, anchor=south west] at ([yshift=1pt]net.north west) {agent network};
\node[font=\tiny, black!65, fill=white, inner sep=1pt, anchor=north]
  at ([yshift=-2pt]scope.south) {one decision scope};
\node[font=\tiny, black!65, fill=white, inner sep=1pt, anchor=north]
  at ([yshift=-11pt]scope.south) {members join and leave};
\node[box, anchor=west, text width=3.15cm] (m1) at (3.5,1.9)
  {\textbf{round 1}\hfill $a_0$: $x_0^{(1)}$\\[1pt]
   \raggedright\scriptsize ``new evidence supports this position''};
\node[box, anchor=west, text width=3.15cm, below=3pt of m1.south west, anchor=north west] (m2)
  {\textbf{round 1}\hfill $a_1$: $x_1^{(1)}$\\[1pt]
   \raggedright\scriptsize ``this constraint rules out that alternative''};
\node[box, anchor=west, text width=3.15cm, below=3pt of m2.south west, anchor=north west] (m3)
  {\textbf{round 3}\hfill $a_4$ left\\[1pt]\raggedright\scriptsize membership update, or silence};
\node[lbl, anchor=south west] at ([yshift=2pt]m1.north west) {on the wire};
\node[font=\tiny, black!65, anchor=north west] at ([yshift=-2pt]m3.south west)
  {positions $+$ explanations $+$ receipts $+$ membership};
\draw[wire] (scope.east) -- ([xshift=-2pt]m2.west);
\node[box, anchor=north west, text width=2.2cm] (state) at (8.85,2.35)
  {\textbf{per member}\\[1pt]\raggedright\scriptsize position in $\mathbb{R}^d$ $\cdot$ facts voiced $\cdot$ facts received};
\node[box, anchor=north west, text width=2.2cm, below=4pt of state.south west] (axes)
  {\textbf{per round}\\[1pt]\raggedright\scriptsize spread $\cdot$ trajectories $\cdot$ momentum $\cdot$ box / hull};
\begin{scope}[shift={(11.75,0.15)}]
  \draw[faint] (0,0) rectangle (2.3,2.3);
  \draw[black!30] (0.2,0.35) -- (0.55,1.7) -- (1.05,0.55) -- cycle;
  \foreach \p in {(0.2,0.35),(0.55,1.7),(1.05,0.55)} {\draw[black!45, -{Straight Barb[length=2pt,round]}] \p -- (1.7,1.65);}
  \draw[black!50, dashed] (1.65,1.6) ellipse (0.35 and 0.27);
  \fill[black!80] (1.95,1.95) circle (1.4pt);
  \node[font=\tiny, black!65, anchor=north east] at (2.25,1.85) {$x^\ast$};
  \node[font=\tiny, black!55, anchor=north] at (0.6,0.28) {start};
\end{scope}
\begin{scope}[on background layer]
  \node[pane, fit={(state)(axes)(11.65,0.05)(14.1,2.6)}, inner sep=7pt] (obs) {};
\end{scope}
\node[lbl, anchor=south west] at ([yshift=1pt]obs.north west) {observer: a monitor};
\draw[wire] (m1.east) -- ++(0.35,0) |- (state.west);
\node[box, draw=black!35, dashed, anchor=north west, text width=2.9cm] (priv) at (-0.55,-0.75)
  {\textbf{not on the wire}\\[1pt]\raggedright\scriptsize what a member worked out and did not send};
\draw[cut] (a2) to[out=-90,in=60] ([xshift=-6pt]priv.north east);
\end{tikzpicture}
\caption{From accumulated communication records to consensus evaluation.}
\label{fig:hero}
\end{figure*}
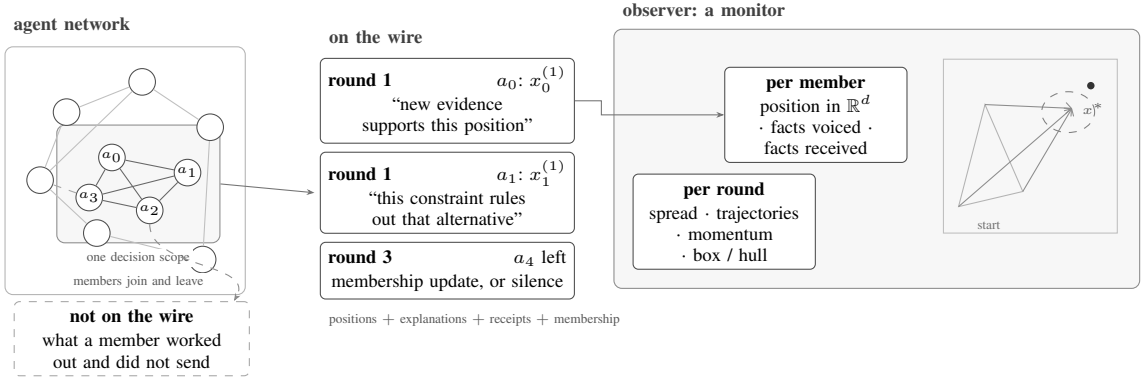

The framework takes agents' communication histories as input and produces measures of consensus
states and trajectories of position changes (\cref{fig:hero}). Its inputs include agents' stated
positions and explanations, delivery records, and membership updates. An observer follows one
decision scope through these records; the accumulated exchange supplies the context for interpreting
changes in the group's positions.
At each stage of an exchange, the observer represents agents' publicly stated positions in a
common, task-defined decision space. For vector-valued tasks, each position lies in $\mathbb{R}^d$,
where the task specifies the dimensions and their meaning. The positions observed at that stage
form the group's distribution. We measure its spread as the mean distance of these positions
to their centroid, describing how closely the group is clustered.

\begin{figure}[t]
\centering
\includegraphics[width=\columnwidth]{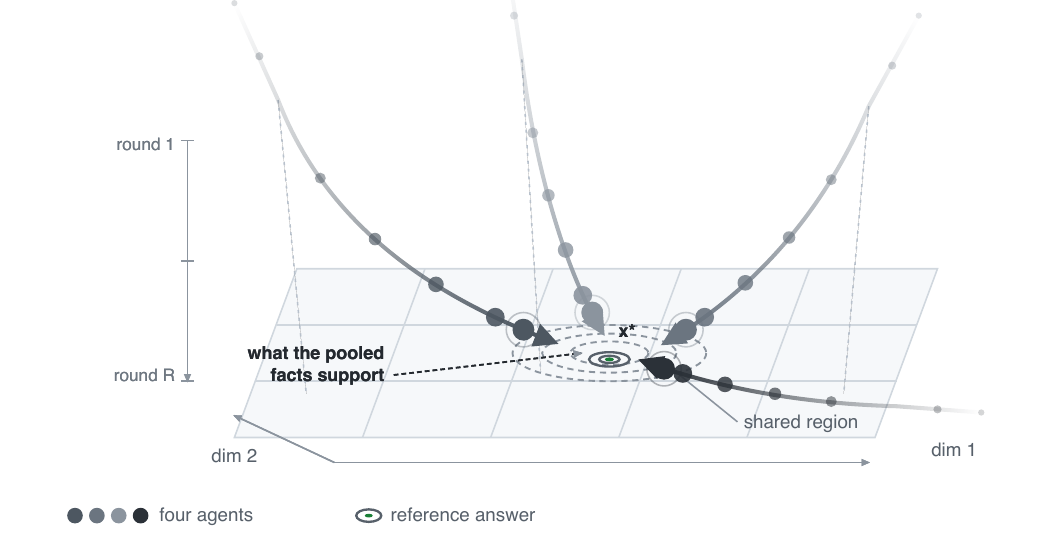}
\caption{Position trajectories approaching a shared region.}
\label{fig:schematic}
\end{figure}

For vector-valued positions, we compare trajectories with two regions defined by the initial
positions: (1) the coordinate-wise box, formed by the minimum and maximum on each dimension;
and (2) the convex hull, containing all convex combinations of the initial positions
\cite{mendes2013multidim,vaidya2013vector,abbas2020centerpoint}. These comparisons show whether
positions remain within or move beyond their starting regions. These initial reference regions
are distinct from the tolerance used to describe a consensus state.
Momentum tracks changes in agents' positions over time and relationships among their trajectories.
\Cref{fig:schematic} illustrates position trajectories approaching a shared region,
with the full-information answer shown as a separate reference. Its perspective view is illustrative;
the framework permits task-defined dimensions.

We relate changes to public information and peers' positions. For pooling, combining facts
whose numerical markers all occur in one public message gives a text-derived target.
We use prior own public messages and peers' messages in the logged prompt, checking hashes
and receipt times. Compatibility means Euclidean distance decreases by more than $10^{-6}$
workers; peer targets are the received peers' mean or coordinate-wise median.
Moves are text-only, peer-only, both, or neither; steps within $10^{-6}$ are unchanged.
Missing observations are separate. These geometric labels are not causal explanations.
Reconciliation text targets remain unassessed: quantity matching cannot resolve duplicates
and cancellations. Round-wise accuracy and agreement track attainment, persistence, and loss.

\section{Case Studies and Experimental Setup}\label{sec:cases}
We use two staffing tasks to examine information pooling and the interpretation of related facts.

\textbf{Information pooling.} We examine the framework through a staffing task with distributed
information. A shared brief specifies the workload and per-worker capacity for three roles,
from which agents compute a baseline staffing vector. Each of four agents receives one private
fact about the following week and adjusts its initial estimate accordingly. Combining all task
facts yields the full-information answer $x^\ast$, used by the evaluator as a scoring benchmark.
For some roles, all facts shift demand in the same direction, placing $x^\ast$ outside the range
of initial estimates; for others, averaging those estimates already gives a close answer.
Each fact contains a distinctive numerical marker, allowing its disclosure to be tracked in
public messages. Agents report staffing estimates in workers per shift, providing three-dimensional
positions whose trajectories can be followed throughout the exchange. Agents are never told $x^\ast$.

\textbf{Fact reconciliation.} The second task uses the same staffing setting but requires agents
to interpret relationships between facts. Six agents hold facts that include two descriptions of
the same contract in different units, and an added contract together with its cancellation.
Computing the full-information benchmark requires counting the duplicate contract once and
applying the cancellation while retaining the contributions of other facts.

\begin{figure*}[t]
\centering
\includegraphics[width=0.95\textwidth]{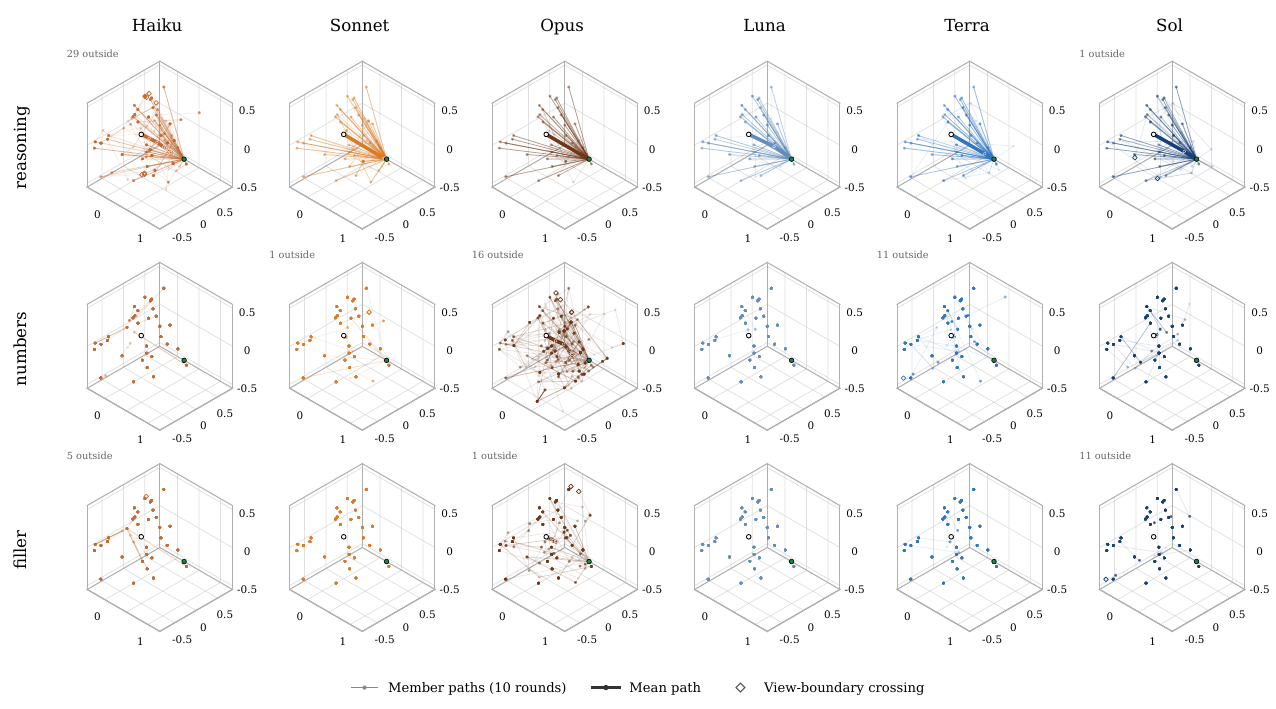}
\caption{Staffing trajectories by model and message condition (eight instances, five repeats; central view).}
\label{fig:tiers}
\end{figure*}

\textbf{Outcome measures.} Progress is measured along the line from the initial group mean (0)
to the full-information answer $x^\ast$ (1); the centroid summarizes the group's trajectory.
The task specifies the coordinate-wise median as the final decision. We record median reach
when this decision is within one worker of $x^\ast$ on every coordinate. The joint accuracy-and-agreement criterion
requires both median reach and a member-to-member range of at most one worker on every coordinate,
using the members present in the final round. This distinguishes
an accurate aggregate decision from accurate decisions accompanied by bounded disagreement.

In \cref{fig:tiers}, instances are rotated toward $x^\ast$ and normalized by the initial
mean-to-answer distance: the starting mean ($\circ$) is 0 and $x^\ast$ ($\bullet$) is 1 on the
first axis. Each panel shows 160 member paths over ten rounds; bold lines average the
normalized centroid paths across eight instances and five repeats. Hollow diamonds mark
view-boundary crossings; 75 of 28,800 recorded positions (0.26\%) lie outside the central view.

\textbf{Message conditions.} We vary the information shared with peers, separately from the
model's reasoning-effort setting. Peers receive each member's vector with its publicly stated
explanation (\emph{reasoning}), the vector alone (\emph{numbers}), or the vector with a fixed
sentence containing no task facts (\emph{filler}). Both tasks compare all three conditions,
with reconciliation filler added in the supplementary experiment. Filler tests whether adding
fixed text without task facts affects agents' responses; it is not length-matched to explanations. Members are told each holds a fact and that
the decision is a median; they are never asked to agree. Numerical positions can encode private
facts through their difference from the shared baseline, so numbers-only retains inferable information.

\textbf{Observation settings.} The \emph{full record} provides all reported positions and actual
arrivals. The \emph{monitor} uses sent messages, delivery acknowledgments, and membership
records. The \emph{transcript-only} setting assumes every message arrived and carries forward
silent members' last positions. These settings distinguish information availability and
membership assumptions. Results use the full record; comparative observer-error evaluation
remains future work.

Our implementation evaluates agents communicating over a Docker network. Each agent runs in
a separate container and accesses its model through an API. Agents exchange messages using
the standard A2A protocol via HTTP, with task data carried in standard data parts and request
signatures provided by a declared extension. A coordinator schedules rounds, while logs retain
each agent's inputs and outputs, delivery acknowledgments, and membership changes for evaluation.

\textbf{Models and repeats.} The design comprises two model families and three tiers: GPT-5.6 Luna,
Terra, and Sol, and Claude Haiku, Sonnet, and Opus 5.5. Each task uses six models, three message
conditions, and eight fixed instances with five repeats, giving 720 runs per task.
Reconciliation adds 160 Sol/Opus explanation/numbers runs and 240 six-model filler runs to the
original 320. Exchanges last ten rounds. Low reasoning effort is requested throughout;
provider routes, resolved models, runtime overrides, and retries remain in the audit records.

Membership experiments vary arrivals, departures, and restarts, including whether changes are
announced or silent. The matrix contains six models, nine conditions, and eight instances,
giving 432 runs with one repeat per cell. Exit conditions distinguish the departing member's
explanations from numbers-only messages. Within-cell repeat variability is not estimated.
Together, the original 1,472 and supplementary 400 runs yield 1,872 valid results after bounded
retries; invalid attempts are excluded and separately retained.

\section{Results}\label{sec:results}

\begin{table}[!t]
\makeatletter
\long\def\@makecaption#1#2{%
  \begingroup\footnotesize\par\centering
  \@IEEEtabletopskipstrut{\normalfont\footnotesize #1.\enspace\textsc{#2}}\par
  \addvspace{0.5\baselineskip}\endgroup
  \@IEEEtablecaptionsepspace}
\makeatother
\centering
\caption{Joint successes / valid runs by model and condition.}
\label{tab:outcomes}
\begingroup
\footnotesize
\setlength{\tabcolsep}{3pt}
\renewcommand{\arraystretch}{1.12}
\begin{tabular*}{\columnwidth}{@{\extracolsep{\fill}}llrrrrrr@{}}
\toprule
& & \multicolumn{3}{c}{Claude 5.5} & \multicolumn{3}{c}{GPT-5.6} \\
\cmidrule(lr){3-5}\cmidrule(l){6-8}
Condition & & Haiku & Sonnet & Opus & Luna & Terra & Sol \\
\midrule
\multicolumn{8}{@{}l}{\textit{Information pooling}} \\
Explanations &  & 21/40 & 39/40 & 40/40 & 40/40 & 40/40 & 40/40 \\
Numbers &  & 0/40 & 0/40 & 8/40 & 0/40 & 0/40 & 0/40 \\
Filler &  & 0/40 & 0/40 & 0/40 & 0/40 & 0/40 & 0/40 \\
\addlinespace[3pt]
\multicolumn{8}{@{}l}{\textit{Fact reconciliation}} \\
Explanations &  & 4/40 & 40/40 & 38/40 & 3/40 & 2/40 & 5/40 \\
Numbers &  & 0/40 & 0/40 & 1/40 & 0/40 & 0/40 & 0/40 \\
Filler &  & 0/40 & 0/40 & 0/40 & 0/40 & 0/40 & 0/40 \\
\addlinespace[3pt]
\multicolumn{8}{@{}l}{\textit{Membership changes}} \\
Late join & A & 4/8 & 6/8 & 6/8 & 6/8 & 7/8 & 7/8 \\
Late join & S & 5/8 & 6/8 & 5/8 & 5/8 & 6/8 & 7/8 \\
Restart & A & 2/8 & 8/8 & 7/8 & 8/8 & 8/8 & 8/8 \\
Restart & S & 3/8 & 7/8 & 4/8 & 8/8 & 8/8 & 8/8 \\
Exit: explanations & S & 3/8 & 8/8 & 8/8 & 8/8 & 7/8 & 7/8 \\
Exit: explanations & A & 0/8 & 7/8 & 6/8 & 5/8 & 7/8 & 7/8 \\
Exit: numbers & S & 0/8 & 5/8 & 7/8 & 3/8 & 4/8 & 1/8 \\
Exit: numbers & A & 0/8 & 3/8 & 7/8 & 1/8 & 3/8 & 0/8 \\
No change & -- & 5/8 & 8/8 & 6/8 & 8/8 & 8/8 & 8/8 \\
\bottomrule
\end{tabular*}
\endgroup

\smallskip
\parbox{\columnwidth}{\footnotesize A: announced; S: silent. Exit labels describe the departing member's messages.}
\medskip
\caption{Evaluation counts (percentages), 240 runs per row.}
\label{tab:metrics}
\begingroup\footnotesize
\setlength{\tabcolsep}{3pt}
\begin{tabular*}{\columnwidth}{@{\extracolsep{\fill}}lrrr@{}}
\toprule
Message & Accurate & Agree & Joint \\
\midrule
\multicolumn{4}{l}{\textit{Pooling}} \\
Explanations & 221\,\textcolor{black!65}{(92.1\%)} & 223\,\textcolor{black!65}{(92.9\%)} & 220\,\textcolor{black!65}{(91.7\%)} \\
Numbers & 8\,\textcolor{black!65}{(3.3\%)} & 21\,\textcolor{black!65}{(8.8\%)} & 8\,\textcolor{black!65}{(3.3\%)} \\
Filler & 1\,\textcolor{black!65}{(0.4\%)} & 0\,\textcolor{black!65}{(0.0\%)} & 0\,\textcolor{black!65}{(0.0\%)} \\
\midrule
\multicolumn{4}{l}{\textit{Reconciliation}} \\
Explanations & 164\,\textcolor{black!65}{(68.3\%)} & 100\,\textcolor{black!65}{(41.7\%)} & 92\,\textcolor{black!65}{(38.3\%)} \\
Numbers & 2\,\textcolor{black!65}{(0.8\%)} & 38\,\textcolor{black!65}{(15.8\%)} & 1\,\textcolor{black!65}{(0.4\%)} \\
Filler & 0\,\textcolor{black!65}{(0.0\%)} & 2\,\textcolor{black!65}{(0.8\%)} & 0\,\textcolor{black!65}{(0.0\%)} \\
\bottomrule
\end{tabular*}\endgroup

\par\vspace{5pt}
\parbox{\columnwidth}{\footnotesize Joint is the intersection of Accurate and Agree; columns are not additive.}
\end{table}

Table~\ref{tab:outcomes} reports totals, not medians of repeat-level success rates:
five eight-instance repeats per task/model cell and one repeat per membership cell.
Explanations are associated with more joint successes in both tasks.

\textbf{Evaluation comparison.} Table~\ref{tab:metrics} applies three evaluations to the same
1,440 task runs at the final round, with a one-worker tolerance per axis.
Reconciliation explanations yielded 72/240 (30.0\%) accurate but disagreeing outcomes;
numbers-only yielded 37/240 (15.4\%) agreeing but inaccurate outcomes. Explanation-condition
joint successes ranged from 2/40 to 40/40 across models. Joint reporting distinguishes
accurate-and-agreeing, accurate-only, agreeing-only, and neither outcomes without establishing
evaluator superiority. Counts are descriptive; repeats reuse fixed instances.

\textbf{Convergence process.} Replaying all 1,440 task runs yields 14,400 round records and
64,800 member transitions. Under explanations, pooling joint successes rise from 88/240
(36.7\%) at round 2 to 220/240 (91.7\%) at round 10; reconciliation rises from 75/240
(31.3\%) to 92/240 (38.3\%). Of 193 reconciliation explanation runs ever accurate, 37 (19.2\%) later
lose accuracy at least once, including recoveries. Pooling explanations yield 7,522/8,640
(87.1\%) unchanged transitions. Among the 1,118 moves, 576 (51.5\%) approach only the
text-derived target, 43 (3.8\%) only a peer target, 475 (42.5\%) both, and 24 (2.1\%)
neither. Final outcomes thus hide intermediate reversals and ambiguous movements.

Groups can agree near the full-information answer even after a member leaves.
With a departing member's explanations visible, joint successes were 41/48 for silent
departures and 32/48 for matched announced departures. This descriptive, single-repeat comparison
pairs the same model and instance.

\section{Discussion}\label{sec:discussion}
Membership determines whose positions enter the assessment; communication history records
information shared over time. A departing member's statements remain in that history, while
a new member's access depends on delivery. Tracking both relates changes in agreement to
changes in available information.

Our tasks use numerical positions and computable full-information references. Open-ended
decisions require a validated mapping from statements to positions. Without reference answers,
quality and informational support need task-specific criteria; agreement can still be assessed
within the chosen representation.

\textbf{Limitations.} Quantity markers do not establish understanding or endorsement;
process labels lack independent semantic validation. Our evaluation assumes trustworthy
logs and membership records. Forged receipts, compromised keys, and fabricated membership
records are outside the validation scope; the experiments do not establish resistance to these attacks.

\section*{AI Use Disclosure}
AI assistants were used to assist code development, data analysis, and manuscript drafting.
The author is responsible for all reported results and citations.
\bibliographystyle{IEEEtran}
\bibliography{refs}
\end{document}